# Stability Study of Photo-inscribed Bragg Gratings in Sapphire Fibers at Very High Temperature

**Antoine BOUREL[1], Léo COLLIARD[1], Thomas BLANCHET[1], Guillaume LAFFONT[1], Sylvain MAGNE[1], Geraud BOUWMANS[2]**

[1] *Université Paris-Saclay, CEA List, F-91120, Palaiseau, France*

[2] *Université de Lille, CNRS, UMR 8523 – PhLAM – Physique des Lasers Atomes et Molécules, F-59000, Lille, France*

antoine.bourel@cea.fr

**ABSTRACT**

Fiber Bragg gratings inscribed by femtosecond laser in sapphire fibers were subjected to thermal cycling up to 1500°C. Adding a multimode silica fiber reduces modal interference, improves the quality of the Bragg spectrum and the measurement accuracy. The measurement uncertainty is about 1.9°C (1σ) at 1500°C.

## 1. INTRODUCTION

Optical fiber sensors are widely used for metrology in harsh environments[1], *e.g.* monitoring at very high temperature[2], high pressure[2] and also in radiative environments[1]. While Fiber Bragg Gratings (FBGs) inscribed in silica fibers survive up to about 1300°C[3], their usage above 700°C to 800°C is limited by the drift of the Bragg wavelength[4]. Sapphire optical fibers are therefore an attractive alternative, thanks to their single-crystal structure and their high melting point (~2040°C)[5]. However, the absence of an optical cladding, resulting in a large core diameter (>75 µm), leads to a highly multimode behavior. As a result, FBGs inscribed in sapphire fibers (S-FBGs) exhibit a broad reflection spectrum (several nanometers) compared to a grating inscribed in a silica fiber. Moreover, the reflection spectrum is altered by a *speckle* pattern, which complicates the precise tracking of the Bragg wavelength. Finally, blackbody radiation adds a further perturbation, especially at high temperature[6]. All these parameters contribute to degrading the measurement uncertainty.

Bragg gratings were inscribed in sapphire fibers using a femtosecond laser and a phase-mask setup. In our study, we focus on a passive method for smoothing the *speckle* phenomenon by adding a multimode fiber to the interrogation setup. The grating was then encapsulated in a sealed alumina capillary and subjected to repeated thermal cycles, made of several plateaus in the [600°C-1500°C] range, during both heating and cooling. This fully passive system thus improves the measurement accuracy, especially at very high temperature.

## 2. INSCRIPTION OF BRAGG GRATINGS IN SAPPHIRE FIBERS

The sapphire fibers used were supplied by Micromaterials (USA), imported, and then connectorized by IDIL (Lannion, France). The end used for interrogation is therefore fitted with an FC/APC connector, and the other end is polished at a 30° angle to minimize the sapphire/air reflection as much as possible. The laser used is a Ti:sapphire femtosecond laser (Astrella, Coherent, USA) emitting 35 fs pulses at a central wavelength of 800 nm and a repetition rate of 1 kHz. The inscription setup consists of a cylindrical lens ($f = 18\,mm$) together with a phase mask (Ibsen Photonics, Denmark), generating a Bragg wavelength in the sapphire fiber around 1545 nm[7]. The S-FBGs were inscribed with pulse energies of 800 µJ for a total inscription duration of 2.5 s. After inscription, the relative amplitude reflected by the grating is approximately 5 dB.

## 3. INTERROGATION SYSTEM OF THE SAPPHIRE BRAGG GRATING

The S-FBG is interrogated using an *Amplified Spontaneous Emission* (ASE) source covering a range from 1520 nm to 1620 nm. The ASE signal is injected into a 2×1 multimode coupler, whose single output is connected to the sapphire fiber, while the other, parallel arm is connected to an optical spectrum analyzer (OSA, AQ6376D, Yokogawa). The sapphire fiber, with a diameter of 100 µm, is strongly multimode; interference between modes therefore produces *speckle* oscillations. To mitigate this effect, a multimode silica fiber with a 105 µm core diameter was inserted in the interrogation path in order to mix the modes and smooth the spectrum[8],[9]. Two parameters were studied: the position of the multimode fiber in the setup and its length. Three configurations were thus tested: after the ASE source (#1), between the coupler and the OSA (#2), and between the coupler and the sapphire fiber (#3) (Fig. 1.a). For each configuration, the fiber length was adjusted up to 500 m (Fig. 1.b). The best compromise between spectral smoothing and insertion loss is obtained when the multimode fiber is placed at position #1, directly after the ASE source, with a fiber length of about 70 m.

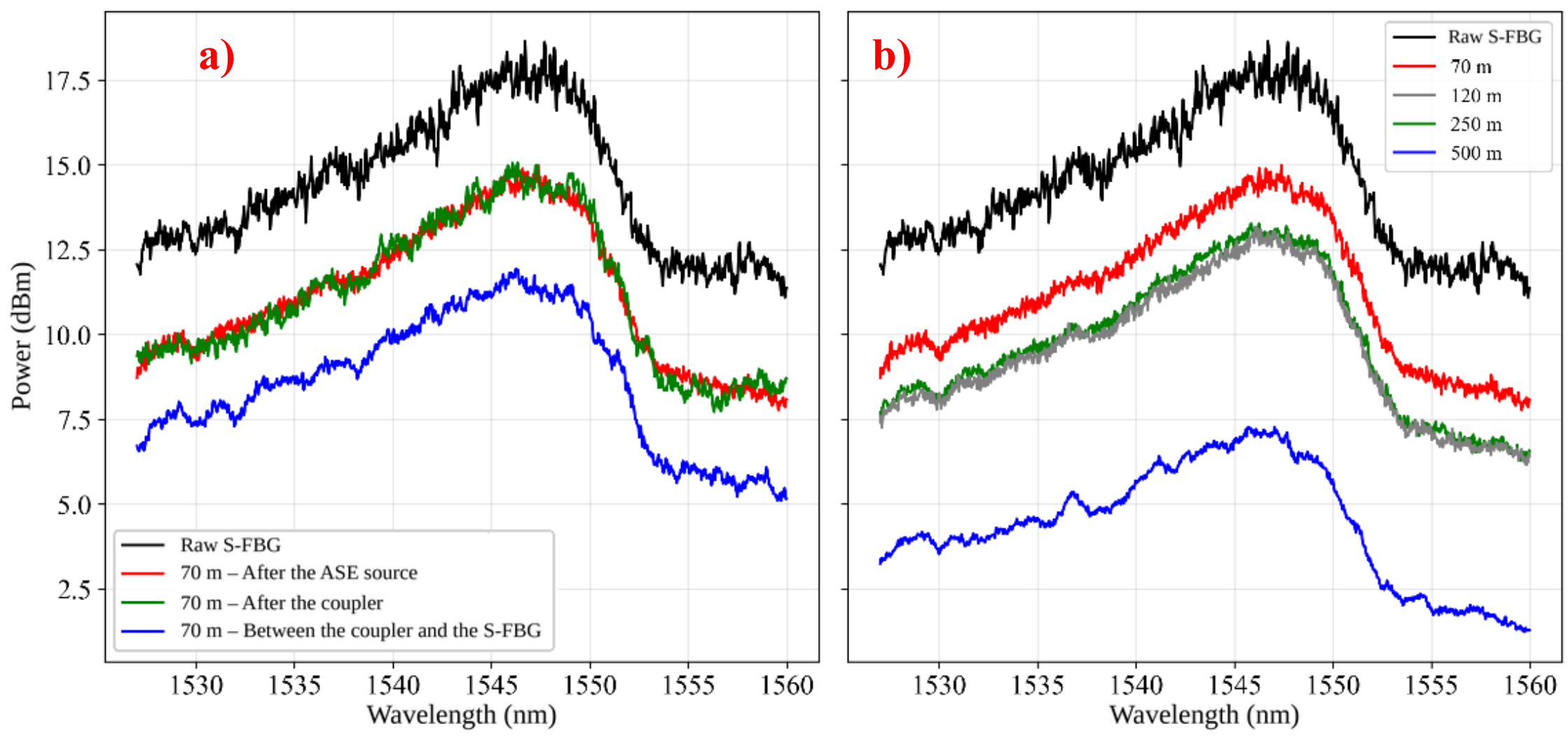

*Figure 1: Reflection spectra of the S-FBG without additional fiber (black) and with additional fiber. (a) 70 m long multimode fiber at positions #1 (red), #2 (green) and #3 (blue), (b) different lengths at position #1: 0 m (black), 70 m (red), 120 m (gray), 250 m (green) and 500 m (blue).*

## 4. THERMAL CYCLES

To avoid any contamination, the grating is encapsulated in an alumina capillary with an inner diameter of 300 µm, sealed at one end with a zirconium-based glue resistant up to 2000°C, while the other end, equipped with the connector, is fixed with a custom-made plug. The fiber is placed horizontally in a tube furnace (Carbolite Gero TF1 16/60/300) reaching 1500°C in air; the temperature is measured close to the grating by a type S thermocouple. The grating underwent six thermal cycles with a 2 h plateau at 1500°C (Fig. 2.a). A moving average is applied to smooth the grating peak, the Bragg wavelength being extracted by the centroid method or by polynomial fitting, both giving consistent results[10]. The temperature is then deduced from the Bragg wavelength shift by applying a 6th-degree polynomial calibration relation (determined during the heating phase).

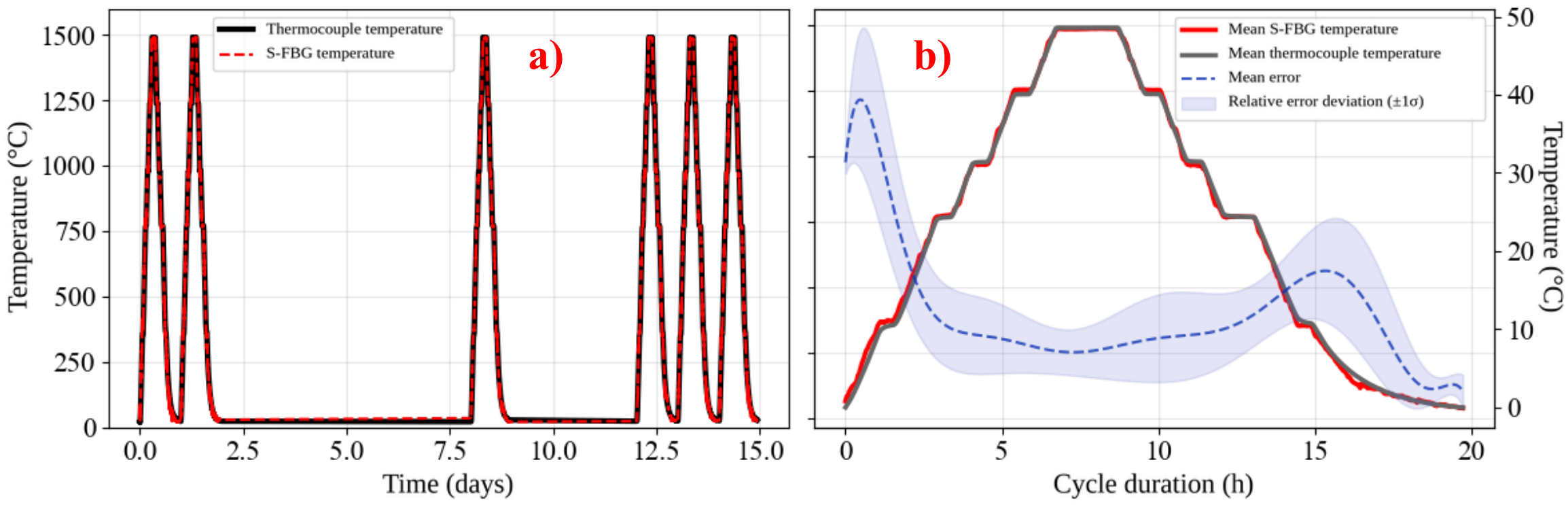

*Figure 2: (a) Temperature measured by the thermocouple (black) and by the S-FBG (red) during the cycles. (b) Mean temperatures per cycle for the thermocouple (black) and the grating (red), with the mean error between the two (blue dotted line) and the light blue band representing its relative deviation ±1σ.*

As shown in Fig. 2.b, stable and smooth temperature plateaus are obtained throughout the cycles, especially thanks to the smoothing of the Bragg spectrum by the multimode fiber. Between 800°C and 1500°C, the difference between the thermocouple and S-FBG measurements is less than 10°C. At 1500°C, the mean observed difference is 6.8 ± 1.5°C, a difference partly attributed to a hysteresis phenomenon[7]. The uncertainty on the peak detection is estimated at 1.9°C (1σ), which reflects an improvement in precision and reproducibility. These results confirm that the proposed approach allows reliable thermal monitoring even at very high temperature, over several cycles.

## 5. Conclusion

Adding a multimode silica fiber to the interrogation setup passively smooths the spectrum of Bragg gratings in sapphire fibers and thus yields reliable temperature measurements up to 1500°C. The study showed that adding a 70 m long smoothing multimode fiber directly after the ASE source offers the best compromise between smoothing and loss of power received by the photodetector. Repeated thermal cycles up to 1500°C showed that, at this temperature, the difference between the thermocouple and the S-FBG remains below 7°C for 2 h, confirming the accuracy and reproducibility of this technique at very high temperature. Experiments lasting several days at high temperature will be carried out for long-term evaluation.

# Etude de la stabilité de réseaux de Bragg photo-inscrits sur fibres Saphir à très haute température

**Antoine BOUREL[1], Léo COLLIARD[1], Thomas BLANCHET[1], Guillaume LAFFONT[1], Sylvain MAGNE[1], Geraud BOUWMANS[2]**

[1] *Université Paris-Saclay, CEA List, F-91120, Palaiseau, France*

[2] *Université de Lille, CNRS, UMR 8523 – PhLAM – Physique des Lasers Atomes et Molécules, F-59000, Lille, France*

antoine.bourel@cea.fr

**RESUME**

Des réseaux de Bragg photo-inscrits par laser femtoseconde dans des fibres en saphir ont été soumis à des cycles thermiques jusqu'à 1500°C. L'ajout d'une fibre multimode silice réduit les interférences modales et améliore la qualité du spectre de Bragg ainsi que la précision de mesure. L'incertitude de mesure est d'environ 1,9°C (1σ) à 1500°C.

## 1. INTRODUCTION

Les capteurs à fibre optique sont très largement utilisés pour la métrologie en environnements difficiles[1], *e.g.* la surveillance à très haute température[2], à forte pression[2] et également en milieu radiatif[1]. Alors que les Réseaux de Bragg (RdBs) photo-inscrits sur les fibres en silice survivent jusqu'à environ 1300°C[3], leur utilisation au-delà de 700°C à 800°C est limitée par la dérive de la longueur d'onde de Bragg[4]. Les fibres optiques en saphir constituent alors une alternative intéressante en vertu de leur structure monocristalline et de leur point de fusion élevé (~2040°C)[5]. Cependant, l'absence de gaine optique résultant en un diamètre de cœur élevé (>75 μm) entraîne un comportement très multimodal. Ainsi, les RdB photo-inscrits dans les fibres en saphir (S-RdB) présentent un spectre de réflexion élargi (plusieurs nanomètres) par comparaison avec celui d'un réseau photo-inscrit sur fibre silice. De plus, le spectre en réflexion est altéré par une figure de *speckle*, compliquant le suivi précis de la longueur d'onde de Bragg. Enfin, le rayonnement du corps noir apporte une perturbation complémentaire, notamment à haute température[6]. Tous ces paramètres contribuent à dégrader l'incertitude de mesure.

Des réseaux de Bragg ont été photo-inscrits dans des fibres en saphir via un laser femtoseconde et un montage par masque de phase. Dans notre étude, nous nous concentrons sur une méthode de lissage passif du phénomène de *speckle* en ajoutant une fibre multimode dans le dispositif d'interrogation. Le réseau a ensuite été encapsulé dans un capillaire en alumine scellé et soumis à des cycles thermiques répétés, constitués de plusieurs plateaux, dans la gamme [600°C-1500°C], en montée et en descente. Ce système totalement passif permet ainsi d'améliorer la précision des mesures, notamment à très haute température.

## 2. PHOTO-INSCRIPTION DE RESEAU DE BRAGG DANS LES FIBRES EN SAPHIR

Les fibres en saphir utilisées ont été fournies par Micromaterials (USA), importées puis connectorisées par IDIL (Lannion, France). L'extrémité utilisée pour l'interrogation est donc munie d'un connecteur FC/APC et l'autre extrémité est polie selon un angle de 30°, pour diminuer au maximum la réflexion saphir/air. Le laser utilisé est un laser femtoseconde Ti:Sa (Astrella, Coherent, USA) émettant des impulsions de 35 fs de durée, à une longueur d'onde centrale de 800 nm, et à un taux de répétition de 1 kHz. Le dispositif d'inscription est composé d'une lentille cylindrique ($f = 18\ mm$) ainsi que d'un masque de phase (Ibsen Photonics, Danemark), générant une longueur d'onde de Bragg dans la fibre en saphir autour de 1545 nm[7]. Les S-RdB ont été photo-inscrits avec des impulsions d'énergie de 800 µJ pendant l'inscription d'une durée totale de 2.5 s. Après inscription, l'amplitude relative réfléchie par le réseau est d'environ 5 dB.

## 3. SYSTEME D'INTERROGATION DU RESEAU DE BRAGG SAPHIR

Le S-RdB est interrogé à l'aide d'une source à émission spontanée amplifiée (*Amplified Spontaneous Emission*, ASE), couvrant une plage allant de 1520 nm à 1620 nm. Le signal ASE est injecté dans un coupleur multimode 2×1, dont la sortie unique est connectée à la fibre en saphir tandis que l'autre bras parallèle est relié à un analyseur de spectre optique (OSA, AQ6376D, Yokogawa). La fibre en saphir d'un diamètre de 100 µm, est fortement multimodale ; les interférences entre modes produisent de ce fait des oscillations de *speckle*. Pour atténuer cet effet, une fibre silice multimode de 105 µm de diamètre de cœur a été insérée sur le chemin d'interrogation afin de mélanger les modes et lisser le spectre[8],[9]. Deux paramètres ont été étudiés : la position dans le montage de la fibre multimode ainsi que sa longueur. Trois configurations furent ainsi testées : après la source ASE (#1), entre le coupleur et l'OSA (#2), et entre le coupleur et la fibre en saphir (#3) (Fig. 1.a). Pour chacune de ces configurations, la longueur de la fibre fut ajustée entre 0 m et 500 m (Fig. 1.b). Ainsi, le meilleur compromis entre le lissage spectral et la perte d'insertion est obtenu lorsque la fibre multimode est placée en position 1, directement après la source ASE, avec une longueur de fibre d'environ 70 m.

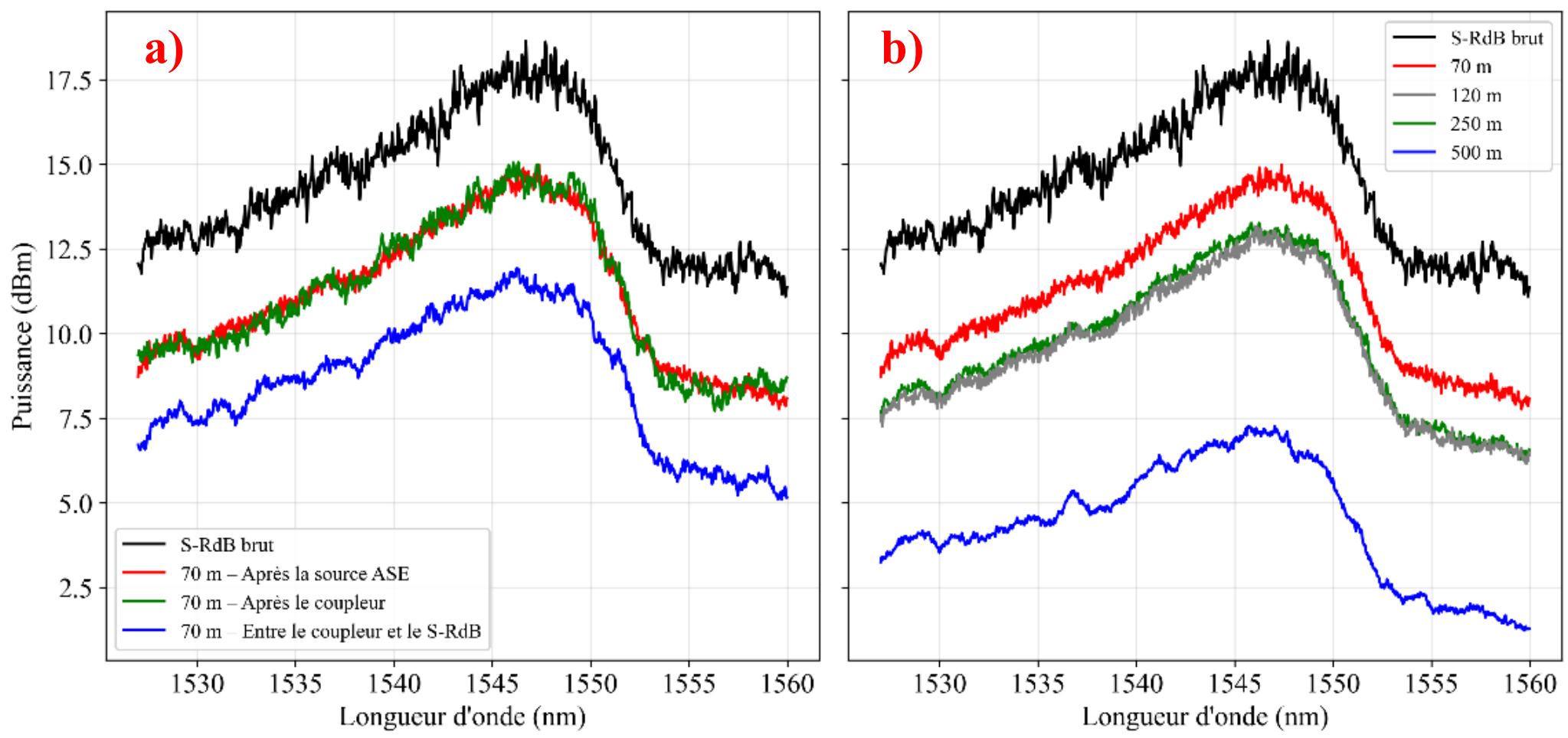

*Figure 1 : Spectres en réflexion du S-RdB sans fibre additionnelle (noir) et avec fibre additionnelle. (a) fibre multimode de 70 m de longueur aux positions #1 (rouge), #2 (vert) et #3 (bleu), (b) différentes longueurs en position #1 : 0 m (noir), 70 m (rouge), 120 m (gris), 250 m (vert) et 500 m (bleu).*

## 4. CYCLES THERMIQUES

Afin d'éviter toute pollution, le réseau est encapsulé dans un capillaire en alumine de diamètre interne 300 µm, scellé à une extrémité avec une colle au zirconium résistante jusqu'à 2000°C, tandis que l'autre extrémité, équipée du connecteur, est fixée à l'aide d'un bouchon sur mesure. La fibre est placée horizontalement dans un four tubulaire (Carbolite Gero TF1 16/60/300) atteignant 1500°C en air ; la température est mesurée à proximité du réseau par un thermocouple de type S. Le réseau a subi six cycles thermiques avec un palier de 2 h à 1500°C (Fig. 2.a). Une moyenne glissante est appliquée pour lisser le pic du réseau, la longueur d'onde de Bragg étant extraite par méthode barycentrique ou ajustement polynomial, les deux donnant des résultats cohérents[10]. La température est enfin déduite du décalage de la longueur d'onde de Bragg en appliquant une relation d'étalonnage sous forme de polynôme de degré 6 (déterminé lors de la phase de montée en température).

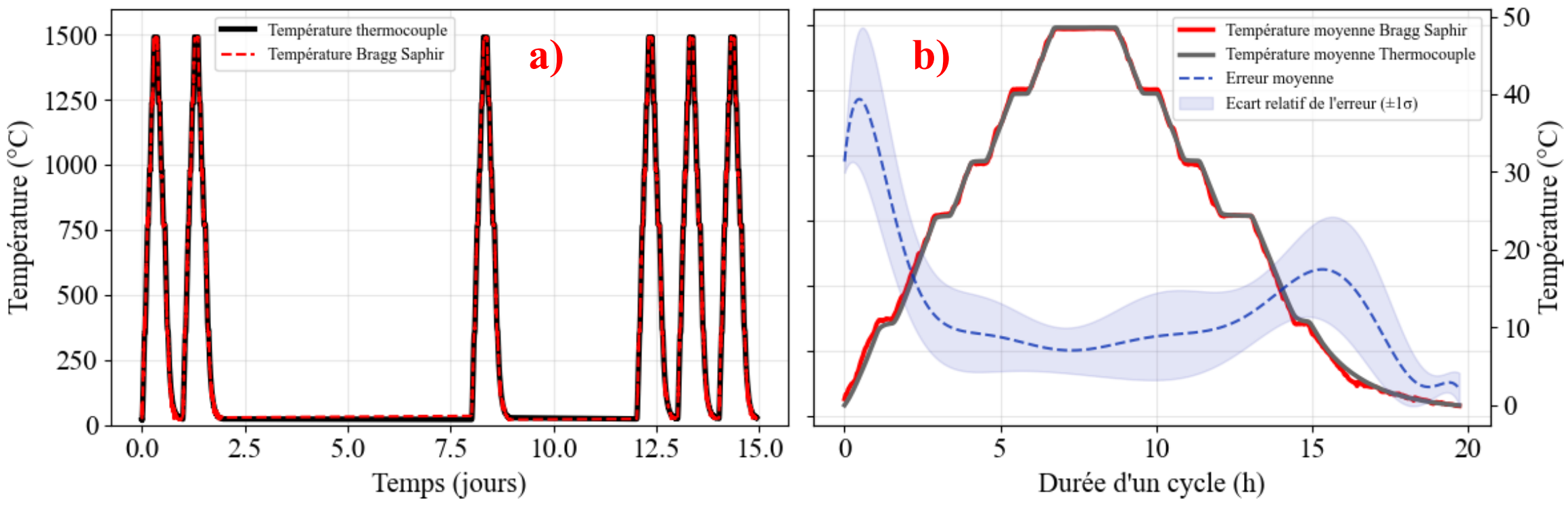

*Figure 2 : (a) Température mesurée par le thermocouple (noir) et par le S-RdB (rouge) lors des cycles. (b) températures moyennes par cycle pour le thermocouple (noir) et le réseau (rouge), avec l'erreur moyenne entre les deux (pointillé bleu) et la bande bleu clair représentant son écart relatif ±1σ.*

Comme illustré en Fig. 2.b, des paliers de température stables et lisses sont obtenus tout au long des cycles, notamment grâce au lissage du spectre de Bragg via la fibre multimode. Entre 800°C et 1500°C, l'écart entre les mesures du thermocouple et du S-RdB est inférieur à 10°C. À 1500°C, l'écart moyen observé est de 6,8 ± 1,5°C, écart en partie attribué à un phénomène d'hystérésis[7]. L'incertitude sur la détection du pic est estimée à 1,9°C (1σ), ce qui traduit une amélioration de la précision et de la reproductibilité. Ces résultats confirment que l'approche proposée permet un suivi thermique fiable même à très haute température et ce, sur plusieurs cycles.

## 5. CONCLUSION

L'ajout d'une fibre multimode en silice dans le dispositif d'interrogation permet de lisser passivement le spectre des réseaux de Bragg sur fibre saphir et ainsi obtenir des mesures de température fiables jusqu'à 1500°C. L'étude a montré que l'ajout d'une fibre multimode de lissage, directement après la source ASE de 70 m de longueur offre le meilleur compromis entre lissage et perte de puissance reçu par le photo-détecteur. Des cycles thermiques répétés jusqu'à 1500°C ont montré qu'à cette température, l'écart entre le thermocouple et le S-RdB reste inférieur à 7°C pendant 2 h, confirmant la précision et la reproductibilité de cette technique à très haute température. Des expériences sur plusieurs jours à haute température seront réalisées pour évaluation à long terme.

## RÉFÉRENCES